\documentstyle[twocolumn,aps,floats]{revtex}

\begin{document}

\draft \tolerance = 10000

\setcounter{topnumber}{1}
\renewcommand{\topfraction}{0.9}
\renewcommand{\textfraction}{0.1}
\renewcommand{\floatpagefraction}{0.9}

%Fixing abstract in twocolumn mode
\twocolumn[\hsize\textwidth\columnwidth\hsize\csname
@twocolumnfalse\endcsname

\title{ The Theory of Fractal Time: Field Equations \\
(the Theory of Almost Inertial Systems  and Modified Lorentz Transformations) }
\author{L.Ya.Kobelev  \\ Department of  Physics, Urals State University \\
 Av. Lenina, 51, Ekaterinburg 620083, Russia \\  E-mail: leonid.kobelev@usu.ru}
\maketitle

\begin{abstract}
Field equations in four order derivatives with respect to time and
space coordinates based on modified classic relativistic energy of the
fractal theory of time and space  are received. It is shown appearing of
new spin characteristics and  new fields with imaginary energies   .

\end{abstract}

 \vspace{1cm}

%Fixing abstract in twocolumn mode
]

\section{Introduction}

In the theory of multifractal time \cite{kob1}-\cite{kob10} for the case
when the fractional dimensions of time $d_{t}$  almost coincide with
integer value (equal unit) ($d_t =1+\sum\beta_{i}L_{i}({\bf r}(t),t)=1+\varepsilon$,
 $|\varepsilon|<<1$)  was shown (see \cite{kob3}) that  for modifying Lorentz
transformations in the theory of fractal time it is necessary to  change factor
$\beta=\sqrt{1-v^{2}/c^{2}}$   by  factor $\beta^{*}=\sqrt[4]{\beta^{4} +4a_{0}^{2}}$.
Then  relativistic energy has form
\begin{equation} \label{1}
E=mc^{2}
\end{equation}
\begin{eqnarray} \label{2}
m=\beta^{*-1}m_{0},  \beta^{2}=1-\frac{v^{2}}{c^{2}}, v\leq{c}
\end{eqnarray}
\begin{eqnarray} \label{3}
m=\beta^{*-1}m_{0}\sqrt{1+\beta^{*2}+
\beta^{2}},\beta^{2}=\frac{v^{2}}{c^{2}}-1, v\geq{c}
\end{eqnarray}
\begin{equation}\label{4}
  p=\beta^{*-1}m_{0}v
\end{equation}
\begin{equation}\label{5}
 \beta^{*}=\sqrt[4]{\beta^{4}+4a_{0}^{2}(t)}
\end{equation}
\begin{equation}\label{6}
  a_{0} =\sum_{i}\beta_{i}F_{0,i}\frac{v}{c}{ct} , \sum_{i} F_{0,i}
  =\sum_{i}\frac{\partial}{\partial {\bf r}}L_{i}
\end{equation}
On the base of fractal  theory in this paper we presented the new equations.
These equations are analogies of relativistic equations (scalar,vector,
tensor, spinor) and valid in the domain of arbitrary velocities including
speed of light if external physical fields don't equal zero (but so small
that fractal dimensions of time is near unit). These equations in the case
of absence of physical fields coincide with the usual relativistic equations,
but give the new fields with imaginary energies (including the case of rest
energies). If the external fields are not equal zero, the received equations
give also a new spin characteristics.

  \section{Equations for  Fields on the Base of Modified Relations for
  Energy}

Let us write formulas (\ref{1})-(\ref{2})  in the form (using the
hypothesis about approximate conservation of the energy-momentum vector in the
space with fractal time \cite{kob3})
\begin{equation}\label{7}
E^{2}  = \frac{E_{0}^{2}}{\beta ^{*2}} =
 \frac{{\bf p}_{0}^{2}c^{2}}{\beta^{*2}} + E_{0}^{2}
\end{equation}
We introduce now the new designation for relativistic energy and momentum similar
to those that have been used in SR ($\beta^{-2}=E^{2}E_{0}^{-2},
E = m_{0}c^{2}\beta^{-2})$ (see (\ref{2})).  Then equation  without roots
has form
\begin{equation}\label{8}
\frac{(E^{2}-{\bf p}^{2}c^{2})^{2}}{(1+4a_{0}^{2}E^{4}E_{0}^{-4})} =
E_{0}^{4}
\end{equation}
\begin{equation}\label{9}
 E=m_{0}\beta ^{-1} , p= m_{0} \text{v}\beta^{-1}
\end{equation}
The equation (\ref{8}) is  the base equation for describing the energy in
the space with multifractal time. For $E$ we receive in the absence of
fields and the momentum equal zero the four solutions:$E=\pm{E_{0}}$,
$E=\pm{iE_{0}}$. Thus, there are particles and anti-particles with real
($E=\pm{E_{0}}$) and imaginary ($E=\pm{iE_{0}}$) masses. The last are new
sort of particles with imaginary masses. These particles are not taxions
because they exist in the domain of velocities  $v\geq0$ (including velocity
$v=c$). For receiving equations for  fields  and particles we use the ordinary
method replacing the energy and the momentum by derivatives. This method
consists in changing the energy $E=E_{0}/(1-v^{2}/c^{2})^{1/2}$ and the
momentum $P=m_{0}v/(1-v^{2}/c^{2})^{1/2}$ by derivatives with respect to time
and space coordinates that used in quantum mechanics: $\hbar=c = 1$,
 $\frac{E}{\beta}$ $\rightarrow i\frac{\partial}{\partial{t}}=\hat{E}$,
$\frac{P_{j}}{\beta} \rightarrow -i\nabla _{j}=\hat{\bf p}$, (j = 1,2,3).
So we obtain ( in case ${v}\leq {c}$) for function   $\Phi({\bf r},t)$ the integral
-differential equation
\begin{equation}\label{10}
  \frac{(\hat{E}^{2}-{\hat{\bf p}}^{2})^{2}}{(1+4a_{0}^{2}\hat{E}^{4}E_{0}^{-4})}
  \Phi({\bf r},t) = E_{0}^{4}\Phi({\bf r},t)
\end{equation}
where $\hat{E}$ and $\hat{p}$  are differential operators and was determined early. For
simplifying this equation we multiply it by operator $(1+4a_{0}^{2})\frac{\hat{E}^{4}}
{E_{0}^{4}})$ (thou it may introduce non-physical solutions but it is makes
the equation (\ref{10})only differential). So we obtain
\begin{eqnarray}\label{11}
(\Box^{2}-4a_{0}^{2}\frac{\partial^{4}}{\partial t^{4}})\Phi({\bf r},t)=
E_{0}^{4}\Phi({\bf r},t)
\end{eqnarray}
where $\Box $ is  D'Alamber operator ($\Box =\Delta
-\frac{\partial^{2}}{\partial t^{2}})$,  $\Delta$  is Laplasian),
$\Phi$  are functions describing particles or fields. For scalar  $\Phi$
equation (\ref{10}) describes the scalar field in the space with fractal
dimensions that originated by the presence of the external physical fields
($a_{0}\neq 0$ .   The corrections in (11) to the usual D'Alamber equation are
the result of modifying the Lorentz transformation. The last is consequences
of fractal nature of time. For receiving the equations in which taken into
account the influence of multifractal structure of time on using the derivatives
and receive  more correct equation  it is  necessary to use the generalized
Riemann-Liouville fractional derivatives (GFD)\cite{kob7},\cite{kob1}. In that
case  equation (11)take the form
\begin{eqnarray}\label{12}
(D_{-,t}^{d_{t}}D_{+,t}^{d_{t}}&-& \Delta )^{2}\Phi({\bf r},t) =
[ E_{0}^{4}+ \nonumber \\
 &+& 4a_{0}^{2}(D_{-,t}^{d_{t}}D_{+,t}^{d_{t}})^{2}]\Phi({\bf r},t)
\end{eqnarray}
where
\begin{equation} \label{13}
D_{+,t}^{d}f(t)=\left( \frac{d}{dt}\right)^{n}\int_{a}^{t}
\frac{f(t^{\prime})dt^{\prime}}{\Gamma
(n-d(t^{\prime}))(t-t^{\prime})^{d(t^{\prime})-n+1}}
\end{equation}
\begin{equation} \label{14}
D_{-,t}^{d}f(t)=\left( \frac{d}{dt}\right)
^{n}\int_{t}^{b}\frac{(-1)^{n}f(t^{\prime})dt^{\prime}}{\Gamma
(n-d(t^{\prime}))(t^{\prime}-t)^{d(t^{\prime})-n+1}}
\end{equation}
where $\Gamma(x)$ is Euler's gamma function, and $a$ and $b$ are some
constants from $[0,\infty)$. In these definitions, as usually, $n=\{d\}+1$
, where $\{d\}$ is the integer part of $d$ if $d\geq 0$ (i.e. $n-1\le
d<n$) and $n=0$ for $d<0$. In this paper we don't consider equations with
fractal derivatives and restrict consideration only by calculation of corrections
from alterations of Lorentz transformation \\

\section{ equations of four and second order in derivatives}

It is useful rewrite (\ref{11}) in the form
\begin{equation}\label{15}
\Box^{2}\Phi({\bf r},t)=(E_{0}^{4}+ 4a_{0}^{2}\frac{\partial^{4}}{\partial
t^{4}})\Phi({\bf r},t)
\end{equation}
Now  introduce  the four component  unit matrix $I$  and the Dirac type matrices
$ \alpha_{i}$: ($\alpha_{i}^{2} =1$, $\alpha_{i} \alpha_{j}+\alpha_{j}\alpha_{i}=0$,
$i\neq j $; $i,j=1,2,3,4 $ ). Than  after usual splitting equation procedure to
the equations (\ref{15})  we receive
\begin{equation}\label{16}
[\Box I + 2a_{0}\frac{\partial ^{2}}{\partial t^{2} }\alpha _{2}]
\Phi  = \alpha _{1} E_{0}^{2}\Phi
\end{equation}
where $ \Phi $ is a four element  bispinor  column
\begin{equation}\label{17}
 \Phi =  \left(
  \begin{array}{c}
    \Phi_{1}\\
   \Phi_{2} \\
    \Phi_{3} \\
    \Phi_{4}
  \end{array}  \right)
\end{equation}
So, we have four equations for $\Phi_{1},...\Phi_{4}$ (\ref{17}).
\begin{equation}\label{18}
  \Box \Phi_{1} - 2a_{0}\frac{\partial ^{2}}{\partial t^{2} }
\Phi_{4}  =  E_{0}^{2}\Phi_{1}
\end{equation}
\begin{equation}\label{19}
  \Box \Phi_{2} - 2a_{0}\frac{\partial ^{2}}{\partial t^{2} }
\Phi_{3}  =  E_{0}^{2}\Phi_{2}
\end{equation}
\begin{equation}\label{20}
  \Box \Phi_{3} - 2a_{0}\frac{\partial ^{2}}{\partial t^{2} }
\Phi_{2}  = - E_{0}^{2}\Phi_{3}
\end{equation}
\begin{equation}\label{21}
  \Box \Phi_{4} - 2a_{0}\frac{\partial ^{2}}{\partial t^{2} }
\Phi_{1}  = - E_{0}^{2}\Phi_{4}
\end{equation}
In the equations (\ref{18})-(\ref{21}) the first two equations describe the
particles or fields with real energies, the last two equations describe the new
particles or fields with imaginary energies.The energies of particles with
real energies depends on behavior of fields with imaginary energies and
vice versa.  The new spin characteristics consequences of these
equations.

\section{equations of first order in derivatives}

For receiving  of first order equations  in derivatives it is necessary
to  introduce the Dirac matrices $ \gamma_{i}$  $(i=0,1,2,3)$  and Dirac
type matrices  $\sigma_{j}$ ($j=0,1,2,3$). These matrices may be used for
splitting of the left-hand side and the right-hand side of each of the
equations (\ref{18})-(\ref{21}). We write the function $a_{0}$ in the form
\begin{eqnarray}\label{22}
  a_{0}=a_{g}+a_{e}+a_{n} &=&
  \beta_{g}L_{g}({\bf r},t)+\beta_{e}L_{e}({\bf
  r},t)+ \nonumber \\  + \beta_{n}L_{n}({\bf r},t)
\end{eqnarray}
where $L_{g}$, $L_{e}$, $L_{n}$ are Lagrangians density of energies for
gravitational, electro-weak and strong fields. Let a module of complex
function $\Phi_{i}$ is the probability to find the particle in a moment $t$
in a point ${\bf r}$. In that case the square root $\sqrt{\Phi_{i}} =
\psi_{i}({\bf r},t)$ gives the function with characteristics: $\psi^*\psi$
has sense of probability. Now if change in equations (\ref{18})-(\ref{21})
the differential operators by $E$ and $P$ , translate in the right-hand side
of equations all members with $a_{0}$ and than   extract a square root from
left-hand side and right-hand side of these equations we obtain (after splitting
we change $E$ and $p$ by ordinary differential operators and use  the
designations $\psi_{j}(i)$ where indexes $i$ corresponds to the indexes $i$
at $\Phi_{i}$ and indexes $j$ corresponds to each splitting component of
$\Phi_{i}$
 \begin{eqnarray}\label{23}
i\gamma_{i}\partial_{i}\sqrt{\Phi_{1}} &=& E_{0}\sigma_{1}\sqrt{\Phi_{1}}+
\sqrt{2a_{g}}\sigma_{2}i\partial_{4}\sqrt{\Phi_{4}} \\
&+& \sqrt{2a_{e}}\sigma_{3}i\partial_{4}\sqrt{\Phi_{4}}+
\sqrt{2a_{n}}\sigma_{4}i\partial_{4}\sqrt{\Phi_{4}} \nonumber
\end{eqnarray}
\begin{eqnarray}\label{24}
  i\gamma_{i}\partial_{i}\sqrt{\Phi_{2}} &=& E_{0}\sigma_{1}\sqrt{\Phi_{2}} +
\sqrt{2a_{g}}\sigma_{2}i\partial_{4}\sqrt{\Phi_{3}}  \\
&+& \sqrt{2a_{e}}\sigma_{3}i\partial_{4}\sqrt{\Phi_{3}} +
\sqrt{2a_{n}}\sigma_{4}i\partial_{4}\sqrt{\Phi_{3}}  \nonumber
\end{eqnarray}
\begin{eqnarray}\label{25}
 i\gamma_{i}\partial_{i}\sqrt{\Phi_{3}}  &=&  E_{0}\sigma_{1}\sqrt{-\Phi_{3}} +
\sqrt{2a_{g}}\sigma_{2}i\partial_{4}\sqrt{\Phi_{2}}  \\
&+& \sqrt{2a_{e}}\sigma_{3}i\partial_{4}\sqrt{\Phi_{2}} +
\sqrt{2a_{n}}\sigma_{4}i\partial_{j}\sqrt{\Phi_{2}}  \nonumber
\end{eqnarray}
\begin{eqnarray}\label{26}
 i\gamma_{i}\partial_{i}\sqrt{\Phi_{4}} &=& E_{0}\sigma_{1}\sqrt{-\Phi_{4}} +
\sqrt{2a_{g}}\sigma_{2}i\partial_{4}\sqrt{\Phi_{1}} \\
&+& \sqrt{2a_{e}}\sigma_{3}i\partial_{4}\sqrt{\Phi_{1}} +
\sqrt{2a_{n}}\sigma_{4}i\partial_{j}\sqrt{\Phi_{1}}  \nonumber
\end{eqnarray}
\begin{equation}\label{27}
\sqrt{\Phi_{j}}= \left(
  \begin{array}{c}
      \psi_{1}(j) \\
      \psi_{2}(j) \\
      \psi_{3}(j) \\
      \psi_{4}(j)
  \end{array}       \right),  (j=1,2,3,4)
\end{equation}
The equations (\ref{23})- (\ref{26}) are generalized Dirac equations
based on  taking into account only the modified by fractal nature of time
Lorentz transformations ( and possibility  of moving with speed of light
as the consequences of it). The eight equations (\ref{23}) and (\ref{24})
describe the particles (or fields) with spin $\frac{\hbar}{2}$ ($\Phi_{i}$
are bispinors), real energy and new characteristics "quasi-spin"
originated by influence of the fields with imaginary energy. The last described
by the equations (\ref{25}) and (\ref{26}). Thus there are two sorts of
particles described by equations (\ref{23})-(\ref{26}): with real energies
and with imaginary energies. Each sort of particles have own anti-particles and
"quasi-spin". In these equations taken into account the influences of all
known  fields (gravitational, electro-weak, strong).\\

\section {Electromagnetic fields with a rest mass (equations Proca)}

Let us consider the case when each of functions $ \Phi_{i} $ is four- vector
of electro-magnetic field ($\Phi_{i}(0),\Phi_{i}(1),\Phi_{i}(2),\Phi_{i}(3)$) with a
rest energy $E_{0}$.  The equations (\ref{15})take form
\begin{equation}\label{28}
\Box \Phi_{\mu}(i) = [\alpha_{1} E_{0}^{2} + 2a_{0}\frac{\partial^{2}}{\partial{t}^{2}}
\alpha_{2}]\Phi_{\mu}(i) ,   (\mu =0,1,2,3)
\end{equation}
So there are four sorts of electro-magnetic Proca fields: fields of Proca
photons and Proca anti-photons  with real and imaginary energies and
different "quasi-spins"  .
For $a_{0}=0)$ all the equations coincide with Proca equations.

 \section{Maxwell equations}

If the Maxwell equations of electro-magnetic field may be considered as the
equations of the Proca in the limits of a rest mass equal zero, in that case
the new sorts of electro-magnetic fields (anti-photons fields and fields with
imaginary energies) as consequences of the Proca fields appear. Let us write
equations for 4-vector electro-magnetic potentials $A_{\mu}(i)$ ( which are
consequences  of equations (\ref{11}) and (\ref{18})-(\ref{21})) for
electro-magnetic fields where  role $E_{0}^{2}$ plays 4-vector of electric
current $j_{{\mu}}$, $({\mu} =0,1,2,3)$
\begin{equation}\label{29}
\Box A_{\mu}(i) = [\alpha_{1}j_{\mu}(i)  + 2a_{0}\frac{\partial^{2}}{\partial{t}^{2}}
\alpha_{2}]A_{\mu}(i) ,   \mu =0,1,2,3
\end{equation}
In equation (\ref{29}) $A_{\mu}(i)$ is a  4-column with respect to $i$.
The  equations (\ref{29}) are generalized Maxwell equations for photon
and anti-photon fields, with usual and "quasi" spins and coincide for
$a_{0}=0$ with usual Maxwell equations. The matrices $\alpha_{1}$ and
$\alpha_{2}$  may be chosen for example as
\begin{equation}\label{30}
  \alpha_{1}=  \left(
 \begin{array}{cccc}
   1 & 0 & 0 & 0 \\
  0 & 1 & 0 & 0 \\
  0 & 0 & -1 & 0 \\
  0 & 0 & 0 & -1
\end{array}          \right),\quad
\alpha_{2}=  \left(
\begin{array}{cccc}
  0 & 0 & 0 & 1 \\
  0 & 0 & 1 & 0 \\
  0 & 1 & 0 & 0 \\
  1 & 0 & 0 & 0
\end{array}            \right)
\end{equation}
The last eight equations (\ref{29}) describe anti-photon fields and its energy
(if electrical charges for anti-photon fields are real) may be imaginary.
In these fields the role of minus-sign charges will play plus-sign charges and
vice versa. There are difference between speed  of photons light  and
anti-photons light. The speeds of photons light  and anti-photons light are equal
only if $a_{0}=0$.

\section{Conclusion}

There are new  consequences of equations based on the modified Lorentz
transformations in the multifractal time theory and now we stress it:
$a$) the existence of two imaginary solutions for the rest energy
($iE_{0}$ and $-iE_{0}$) and thus the existence of new class of particles
with imaginary mass. Because its velocities are arbitrary ( $0\leq v \leq \infty$
they are not taxions );  $b$) the appearance of  new "quasi-spin"
characteristics. The cause of it lays  in the more high order ( four order)
differential equations then usual equations and in additional splitting of
square roots and taking into account the existence of external fields
($a_{0}$). The physical sense and physical nature of new spin characteristics
are not clear. The nature of the additional "spins" originated by  additional
decompositions of square roots for equations of four order and physical sense
of it needs in special  investigation; ;  $c$) all equations coincide with
known physical equation if fractal dimension of time is integer ( so in that
sense the theory is not contradicts known physical theories). In the fractal
theory of time the last case corresponds to vanishing of physical fields
(the last originates the fractional dimensions of time); $d$) the theory
use the improved classical relations for relativistic energy which take
into attention the fractional dimensions of time and allow motions with
arbitrary velocities (including velocities equal the speed of light);
e) it may be shown that the equations for case $v\geq{c}$ some differs at
equations used in the paper but   main results the theory based on them
coincide qualitative with the results  used in this paper(if use the equations
of this paper for case $v\geq{c}$ ). \\
The presented in this paper the theory  based on a relative motions in
almost inertial  systems which in turn based on the multifractal time theory
\cite{kob1} and gives the new describing for  characteristics of moving
bodies (energy, momentum, mass and so on). The main results of this theory
used in this paper are: $a$) the possibility of moving with arbitrary
velocities without appearance of infinitum energy and imaginary mass;
$b$) existence of maximum energy if $v=c$;  $c$) possibility of experimental
verification  the main results of the theory.\\
The theory \cite{kob1}-\cite{kob10} describes the Universe as an open systems
(the theory of open systems see in \cite{klim}). This theory coincides with SR
after transition to inertial systems (if neglect by the fractional dimensions
of time) or almost coincides (the differences are non-essential) for
velocities $v<c$. The movement of bodies with velocities that  exceed the
speed of light is accompanied by a series of physical effect's which can
be found  by experiments (these effects was considered in the separate
papers (\cite{kob3}, \cite{kob5}, \cite{kob9}) in more details). It is
shown in these papers the necessity to receive the particles with energies
$\sim E_{0}$ $10^{3}$ $1/\sqrt{t}$  and  $\frac{\partial}{\partial x} a_{e}<
\frac{\partial}{\partial x} a_{g}$ for verification of the theory.If accelerate
the particles by electric fields   then  $ E_{0}[2\tilde{E}(Mc_{2})^{-1}t ]^{-\frac{1}{2}}
< E_{0}(\sqrt{2a_{g}t})^{-1})$. In this formula $\tilde{E} $ is the
electric field strength, $ M $ is the mass of electric charges originated
the $\tilde{E}$ .  It is useful to pay attention to the problem of receiving
the particles with such energies of the physicians  of known accelerate centers.
If  the possibilities for organizing such works will be found, the results are useful
for development of our views on the nature of an energy, the  time and the space.

\end{document}